**Polarization dependence of double resonant Raman scattering band in bilayer graphene**

Jae-Ung Lee,[a, 1] Ngor Mbaye Seck,[a, 1] Duhee Yoon,[a, 2] Seon-Myeong Choi,[b] Young-Woo Son,[b] and Hyeonsik Cheong[a, *]

[a]*Department of Physics, Sogang University, Seoul 121-742, Korea*

[b]*School of Computational Sciences, Korean Institute for Advanced Study, Seoul 130-722, Korea*

**Footnote**

* Corresponding author. E-mail address: hcheong@sogang.ac.kr (H. Cheong)

[1] These authors contributed equally.

[2] Present address: Engineering Department, University of Cambridge, Cambridge CB3 0FA, United Kingdom

**Abstract**

The polarization dependence of the double resonant Raman scattering (2D) band in bilayer graphene (BLG) is studied as a function of the excitation laser energy. It has been known that the complex shape of the 2D band of BLG can be decomposed into four Lorentzian peaks with different Raman frequency shifts attributable to four individual scattering paths in the energy-momentum space. From our polarization dependence study, however, we reveal that each of the four different peaks is actually doubly degenerate in its scattering channels, i.e., two different scattering paths with similar Raman frequency shifts for each peak. We find theoretically that one of these two paths, ignored for a long time, has a small contribution to their scattering intensities but are critical in understanding their polarization dependences. Because of this, the maximum-to-minimum intensity ratios of the four peaks show a strong dependence on the excitation energy, unlike the case of single-layer graphene (SLG). Our findings thus reveal another interesting aspect of electron-phonon interactions in graphitic systems.

1. Introduction

Bilayer graphene is attracting much interest owing to the possibility of modifying its electronic band structure and other physical properties using external perturbations. The possibility of opening a band gap using an external electric field[1–4] or strain[5] is an important issue for applications of graphene in electronic and optoelectronic devices. In contrast to single-layer graphene, bilayer graphene has massive chiral quasi-particles with a parabolic energy dispersion at low energy,[6–8] and there are two conduction and two valence bands, split due to the interlayer coupling. Raman spectroscopy has been widely used in graphene research to investigate the properties of single- and few-layer graphene.[9] In addition to determination of the number of layers,[10,11] Raman spectroscopy can be used to estimate carrier density,[12–18] strain[19–22] and temperature[23] of a given graphene sample.

The 2D band of the Raman spectrum of graphene is of particular interest. It results from emission of 2 phonons near the K point of the Brillouin zone through a double-resonance Raman scattering process. Because of the resonance conditions, the frequency and the line shape of this band reflect both the phonon dispersion and the electronic band structure. As a result, this band is dispersive[24]: its position and line shape depend on the excitation wavelength because the resonance conditions depend on the incoming photon energy. Therefore, the details of the electronic band structure can be revealed through careful examination of the Raman 2D band of bilayer graphene. The 2D band of bilayer graphene is commonly decomposed into 4 Lorentzian peaks, corresponding to 4 scattering pathways.[12] Each of these 4 Lorentzian peaks contains information on the details of electron-photon and electron-phonon interactions involved in the Raman scattering process. In *single*-layer graphene, a strong polarization dependence of the 2D intensity was observed and interpreted[25] in terms of anisotropic electron-photon and electron-phonon interactions[26,27] reflecting the peculiar nature of Dirac electrons. In this work, we investigated the polarization dependence of the 2D band of *bilayer* graphene and its excitation energy dependence. We have found that the polarization dependence is qualitatively different from that of single-layer graphene. This difference can be explained as being due to the interlayer coupling in bilayer graphene.

**2. Methods**

2.1 Sample preparation

We used 4 different bilayer graphene samples for the measurements: three samples prepared on Si substrates covered with a 100 nm-thick $SiO_2$ layer and one on a 300 nm-thick $SiO_2$ layer. For comparison, we also performed the measurements for a single-layer graphene sample on a 300-nm $SiO_2$/Si substrate. The samples on Si substrates with a 100 nm-thick $SiO_2$ layer are more appropriate for investigating the excitation energy dependence due to the interference effect from the substrate.[28] All of the samples are prepared directly on the substrate by mechanical exfoliation from natural graphite flakes. The number of layers was identified by the line shape of the Raman 2D band.[10,11] Although there are small differences in the Raman features due to different level of doping from the environment[17,18] and residual strain,[29] the observed polarization dependence is robust. For the analysis in bilayer graphene, we used averaged values from the 4 samples to reduce experimental uncertainties.

2.2 Polarized Raman measurements

For the polarized Raman measurements, 4 different excitation sources were used: the 441.6-nm (2.81 eV) line of a He-Cd laser, the 488-nm (2.54 eV) and 514.5-nm (2.41 eV) lines of an Ar ion laser, and the 632.8-nm (1.96 eV) line of a He-Ne laser. The laser beam was focused onto the sample by a 50× microscope objective lens (0.8 N.A.), and the scattered light was collected and collimated by the same objective. An analyzer was placed after the objective to select the polarization of the scattered photons. For the 441.6-, 488- and 514.5-nm excitation, the scattered signal was dispersed with a Jobin-Yvon Triax 550 spectrometer (1200 grooves/mm) and detected with a liquid-nitrogen-cooled back-illuminated charge-coupled-device (CCD) detector. For the 632.8-nm excitation, the scattered signal was dispersed with a Jobin-Yvon Triax 320 spectrometer (1200 grooves/mm) and detected with a thermoelectrically cooled, back-Illuminated, deep-depletion CCD detector. The laser power was kept below 1 mW to avoid local heating. The incident laser polarization was fixed, and the spectra were measured as a function of the analyzer angle. An achromatic half-wave plate was placed in front of the spectrometer entrance slit and adjusted to keep the polarization direction of the signal entering the spectrometer constant with respect to the groove direction of the grating regardless of the analyzer angle.

2.3 *Ab initio* calculation details

In order to calculate the 2D peak positions and the dipole transition matrix elements, we performed calculations on electronic band structures of BLG based on the first-principles self-consistent pseudopotential method[30] using the local density approximation for exchange-correlation functional and GW correction,[31–33] and on their phonon dispersions by using the density-functional perturbation theory.[30] The ion core of carbon atoms is described by a norm-conserving pseudopotential.[34] A k-point sampling of the $48\times 48\times 1$ grid uniformly distributed in the two-dimensional Brillouin zone is used in self-consistent calculations and a $6\times 6\times 1$ grid is used to calculate the dynamical matrices.

### 3. Results and discussions

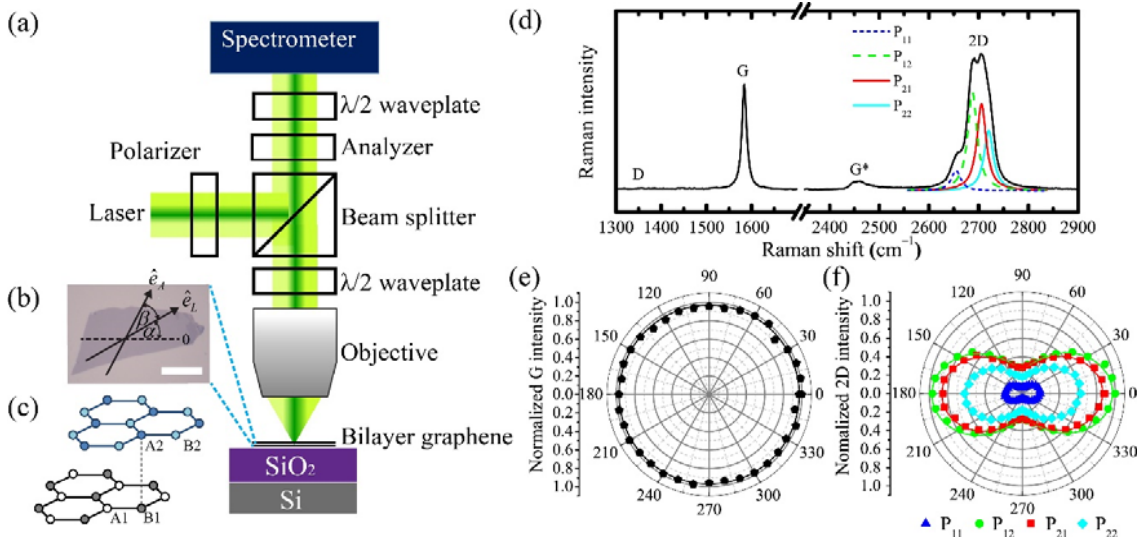

**Fig. 1 -** (a) Schematic diagrams of polarized Raman measurements, (b) optical microscope image of the sample and the definition of the incident polarization angle (*α*) and the analyzer angle (*β*) The scale bar is 20 μm. (c) Atomic structure of Bernal-stacked bilayer graphene. (d) Raman spectrum of bilayer graphene for the excitation energy of 2.41 eV (514.5 nm). 4 single Lorentzian components of the 2D band obtained from fitting are also shown. $P_{11}$, *etc*. refer to the 4 scattering paths defined in Fig. 4. Polarization dependence of the intensity of Raman (e) G band and (f) each component of the 2D band as a function of the analyzer angle.

A schematic diagram for the polarized Raman measurements is shown in Fig. 1(a), and Fig. 1(b) is an optical microscope image of a Bernal-stacked [Fig. 1(c)] bilayer graphene sample. Figure 1(d) shows a typical Raman spectrum measured with the 514.5-nm (2.41eV) excitation. There is no indication of the D peak, which attests to the high quality of the sample. The 2D band has a typical line shape of Bernal-stacked bilayer graphene.[10,11] It can be fitted with 4 Lorentzian peaks as shown in Fig. 1(d), corresponding to the 4 'allowed' scattering paths.

Fig. 1(e) and (f) show the intensities of the G band and each Lorentzian component of the 2D band as functions of the analyzer angle [$\beta$, see Fig. 1(b)]. Because the G peak is generated by degenerated $E_g$ phonon modes whose Raman tensors have the same symmetry as the $E_{2g}$ phonons in single-layer graphene,[35] the G peak of bilayer graphene is isotropic as that of single-layer graphene.[25] On the other hand, Fig. 1(f) shows that each of the 4 Lorentzian peaks of the 2D band has a strong polarization dependence as in single-layer graphene.[25,36] It seems that the overall polarization dependence of the Raman signal is not much different from that of single-layer graphene. It means that the effects of the interlayer interaction on the polarization dependence of the electron-photon interactions and on the electron-phonon interactions are small.[37]

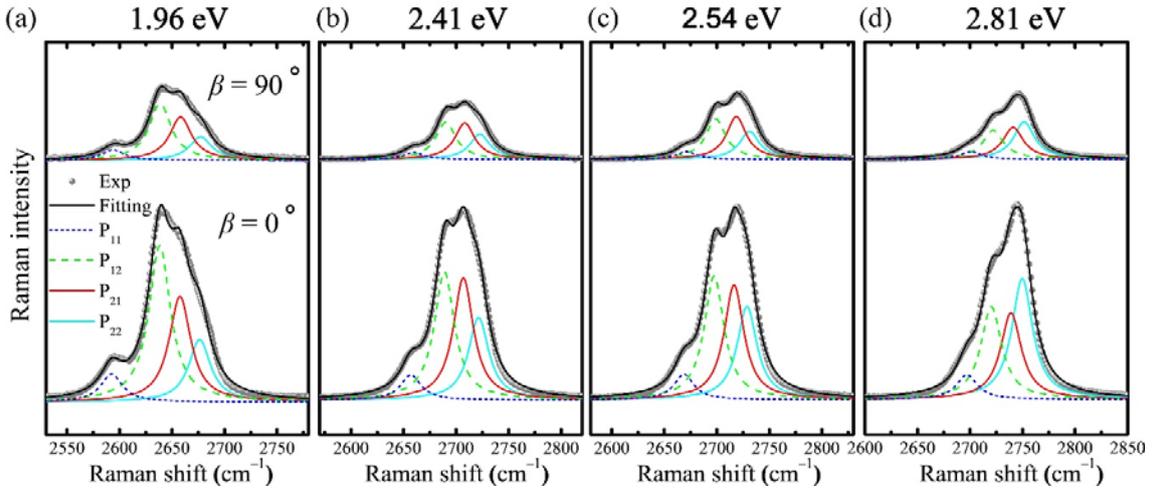

**Fig. 2** - Raman spectrum of the 2D band of bilayer graphene for $\beta=0°$ (lower) and for $\beta=90°$ (upper) with 4 different excitation energies: (a) 1.96 eV (632.8 nm), (b) 2.41 eV (514.5 nm), (c) 2.54 eV (488 nm), and (d) 2.81 eV (441.6 nm).

For further investigation, we measured the Raman spectrum using different excitation energies. Fig. 2 shows the Raman 2D band measured with 4 different excitation energies. The spectra taken with the analyzer angle [$\hat{e}_A$, see Fig. 1(b)] parallel ($\beta=0°$) and perpendicular ($\beta=90°$) to the incident polarization direction ($\hat{e}_L$) are compared. The frequency of the Raman 2D band depends on the excitation energy,[24] because the momentum of the phonon involved in the double-resonance Raman process depends on the excitation energy. The position and the width of the 2D band varies slightly between samples, probably due to different degrees of unintentional doping[17,18] and residual strain.[29] The dispersion of the peak positions agrees with previous results which concluded that the so-called inner process is dominant.[38] One can also observe that the relative intensities of the 4 peaks change with the excitation. This is due to the different Raman scattering cross-section and phonon dynamics, as recently reported by Mafra *et al.*[39]

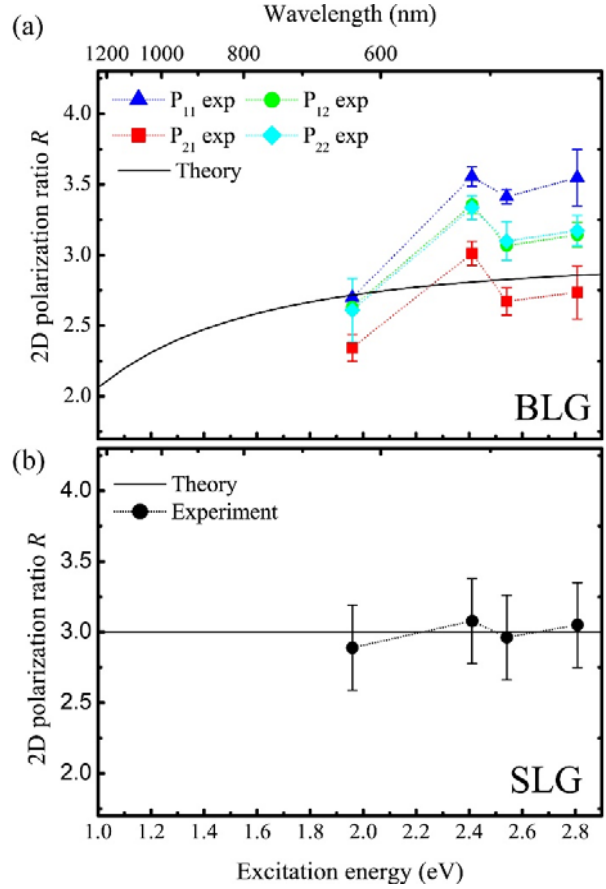

**Fig. 3 -** The polarization ratio *R* of bi- (upper) and single-layer (lower) graphene. The black solid curves are from model calculations.

Upon close inspection, we further observe that the relative intensities of the 4 peaks are slightly different in the two polarizations even for the same excitation energy. We define the polarization ratio *R* as

$$R = I_{//} / I_{\perp},$$

where $I_{//}$ is the Raman intensity when the analyzer angle is parallel to the incident polarization ($\beta=0°$) and $I_{\perp}$ is that when the analyzer and the incident polarization are perpendicular to each other ($\beta=90°$). Figure 3(a) shows the $R$ values for each of the 4 components of the 2D band as a function of the excitation energy. The $R$ values are different for the 4 components at any excitation energy. Furthermore, the $R$ value exhibits excitation energy dependence, showing a smaller value for the lowest excitation energy. In comparison, the $R$ value for the single-layer sample shows no dependence on the excitation energy as shown in Fig. 3(b). For single-layer graphene, an $R$ value of ~3 can be obtained from a simple heuristic model.[25] If the interlayer coupling in bilayer graphene can be ignored, the polarization ratio $R$ of the 2D band of bilayer graphene should be similar to that of single-layer graphene. However, the data show clear departure from this simple expectation. The difference in the $R$ values of the 4 components for a given excitation energy may be understood as being due to different electron-photon and electron-phonon interactions in the scattering process for each peak. The excitation energy dependence of the $R$ value for a given peak component, on the other hand, requires a close inspection of the scattering processes.

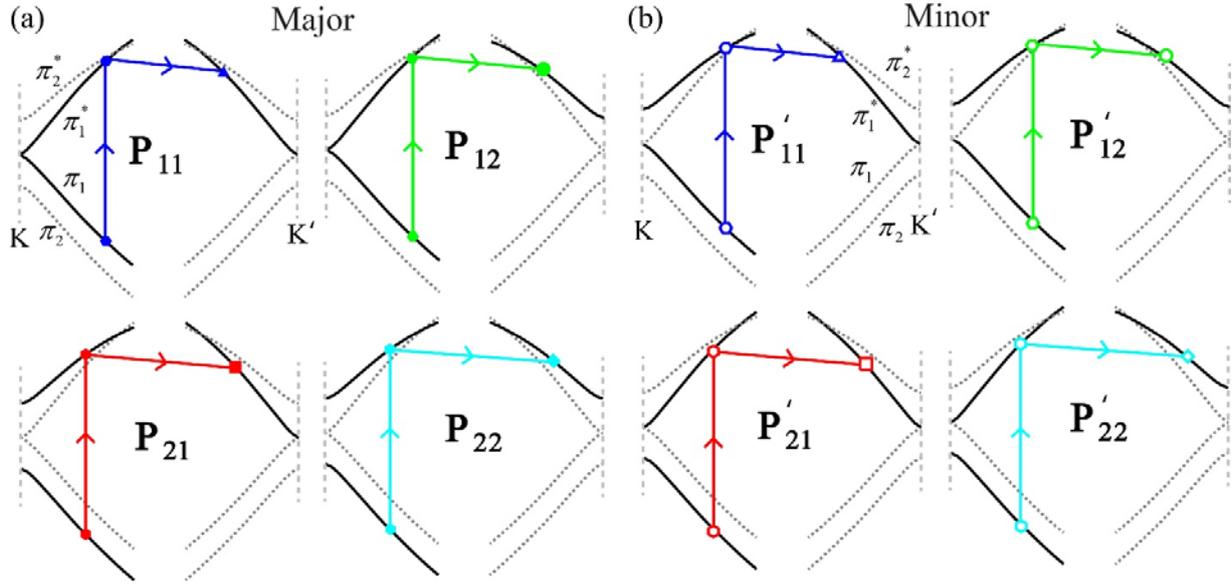

**Fig. 4** - Scattering processes of the Raman 2D band of bilayer graphene with excitation energy 2.41 eV (514.5 nm) correspond to each peaks $P_{11}$, $P_{12}$, $P_{21}$ and $P_{22}$ for (a) major and (b) minor transitions.

The scattering processes of the Raman 2D band for Bernal-stacked bilayer graphene are shown in Fig. 4. Bilayer graphene has 4 electronic bands consisting of 2 conduction ($\pi_1^*$ and $\pi_2^*$) and 2 valence bands ($\pi_1$ and $\pi_2$). A process labeled as $P_{ij}$ means that an electron near the K point is excited from the valence band $\pi_i$ to the conduction band $\pi_i^*$ ($i$=1,2), and then scatters to the $\pi_j^*$ band ($j$=1,2) near the K′ point by emitting a phonon. The electron returns to the same position in the reciprocal space by emitting another phonon with the same momentum and emits a Raman scattered photon. In most analyses of the 2D band of bilayer graphene, only these four 'major' pathways shown in Fig. 4(a), corresponding to the 4 components of the 2D band, are considered.[40] Several recent studies[20,21,38,41–43] have established that the most dominant scattering path for the 2D band is the so-called 'inner process', in which electrons on the K-M-K´ high-symmetry direction give the dominant contribution to the 2D band. Fig. 4(b) shows four additional 'minor' scattering pathways which includes the optical transitions between $\pi_i$ and $\pi_j^*$ for $i \neq j$. These minor pathways have been largely ignored in the analysis of the 2D band based on symmetry arguments.[40] Since the minor processes have small intensities and their Raman shifts are very similar to those of the corresponding major processes [See Fig. 5(a)], the 4-Lorentzian-peak fitting produces excellent results. However, the major and minor processes have different polarization dependences which result in peculiarities in the overall polarization dependence of the 2D band as will be discussed below.

For theoretical analysis of the polarization dependence of the major and the minor processes, we consider the Hamiltonian of Bernal-stacked bilayer graphene which can be written as

$$H_{\vec{p}} = \sum_\alpha \pi_{\vec{p}} c_{B\alpha}^\dagger(\vec{p}) c_{A\alpha}(\vec{p}) - \gamma_1 c_{A1}^\dagger(\vec{p}) c_{B2}(\vec{p}) + (\text{h.c.}),$$

where $c_{A\alpha}$ ($c_{B\alpha}^\dagger$) is an annihilation (creation) operator for an electron at site $A\alpha$ ($B\alpha$) in the lower ($\alpha=1$) and upper layers [$\alpha=2$, see Fig. 1(c)]. $\gamma_1$ is the nearest interlayer hopping (between A2 and B1 carbons); and $\pi_{\vec{p}} = v_0(p_x + ip_y)$, where $\vec{p} = (p_x, p_y)$ is the crystal momentum from K point, and $v_0$ is the Fermi velocity.[44–46]

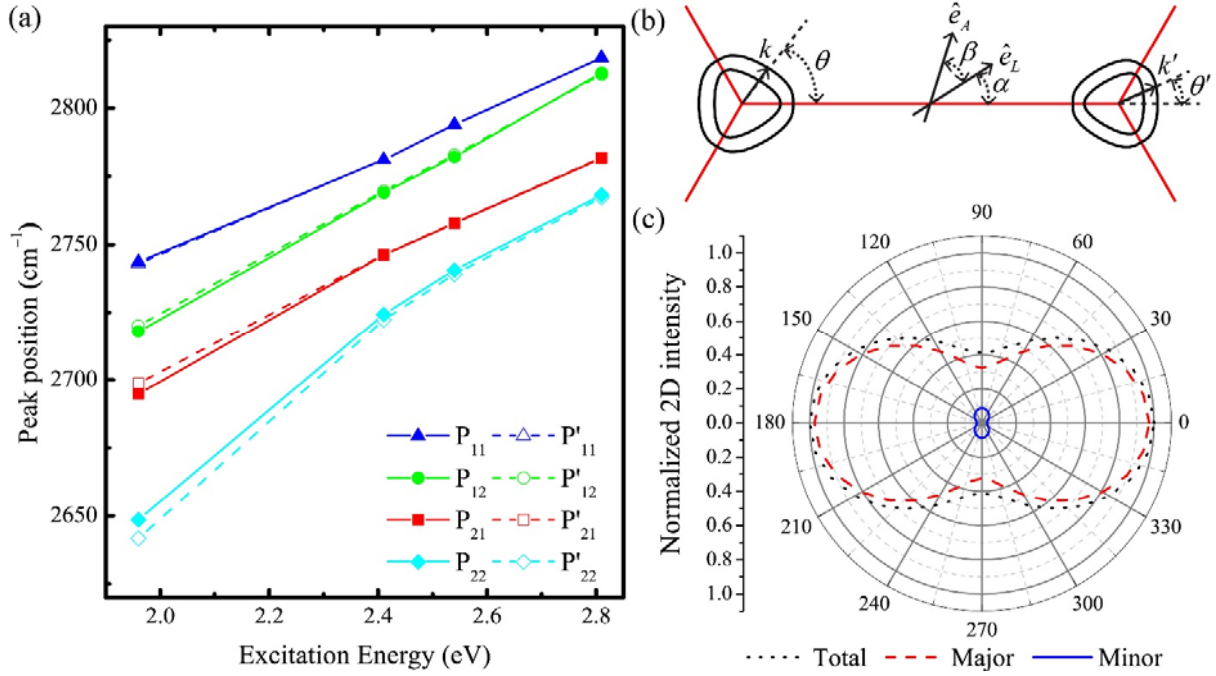

**Fig. 5 -** (a) Calculated peak positions from the major and minor processes from DFT calculations. (b) Definition of the angles: incident laser polarization direction ($\alpha$), the analyzer direction ($\beta$), and the angles of the wavevectors $\vec{p}\,(\theta)$ and $\vec{p}\,'(\theta')$. (c) The polarization dependence of the major (red), minor (blue), and total (black) 2D Raman intensities for 1.96 eV excitation energy.

If we consider the electron-photon interactions for a laser energy of ~2 eV, for example, the eigenenergies of the states involved in the Raman scattering are much larger than $\gamma_1$, and so the interlayer interaction term can be treated as a perturbation. Therefore, the four (unnormalized) eigenstates of the Hamiltonian $H_{\vec{p}}$ in a basis $\Psi = (\psi_{A1}, \psi_{B1}, \psi_{A2}, \psi_{B2})$ are approximated by

$$\begin{pmatrix} e^{-i\theta} \\ \frac{\gamma_1}{2pv_0}-1 \\ \frac{\gamma_1}{2pv_0}-1 \\ e^{i\theta} \end{pmatrix}, \begin{pmatrix} -e^{-i\theta} \\ \frac{\gamma_1}{2pv_0}+1 \\ \frac{-\gamma_1}{2pv_0}-1 \\ e^{i\theta} \end{pmatrix}, \begin{pmatrix} -e^{-i\theta} \\ \frac{\gamma_1}{2pv_0}-1 \\ \frac{-\gamma_1}{2pv_0}+1 \\ e^{i\theta} \end{pmatrix}, \begin{pmatrix} e^{-i\theta} \\ \frac{\gamma_1}{2pv_0}+1 \\ \frac{\gamma_1}{2pv_0}+1 \\ e^{i\theta} \end{pmatrix}, \quad (1)$$

for four eigenenergies, $\varepsilon_{\pi_1} \cong -v_0 p + \frac{1}{2}\gamma_1$, $\varepsilon_{\pi_2} \cong -v_0 p - \frac{1}{2}\gamma_1$, $\varepsilon_{\pi_1^*} \cong v_0 p - \frac{1}{2}\gamma_1$, and $\varepsilon_{\pi_2^*} \cong v_0 p + \frac{1}{2}\gamma_1$, respectively, where $\theta$ is an angle between the momenta of electron and the KΓ direction of the first Brillouin zone of graphene [see Fig. 5(b)]. Around that energy range, the energy bands have linear dispersions like single-layer graphene except for $\pm\frac{1}{2}\gamma_1$ energy shifts, and the (0, $\frac{\gamma_1}{2pv_0}$, $\pm\frac{\gamma_1}{2pv_0}$, 0) terms in the eigenstates are added to that of single-layer graphene. If we do not consider the interlayer interaction ($\gamma_1 = 0$), minor transitions ($\pi_1 \to \pi_2^*$, $\pi_2 \to \pi_1^*$) are forbidden. But, if we consider the perturbation term, the optical transition is possible, and the transition amplitude depends on the incident laser energy.

By using the dipole approximation, we find that the polarization dependence of the optical transition matrix element of a minor transition is opposite to that of a major transition. When the incident (emission) light is polarized along $\hat{e}_i = (\cos\alpha, \sin\alpha, 0)$ [$\hat{e}_s = (\cos\alpha', \sin\alpha', 0)$] [see Fig. 5(b)], the transition matrix element of a major transition at $\vec{p} = p(\cos\theta, \sin\theta)$ [$\vec{p}' = p'(\cos\theta', \sin\theta')$] is proportional to $M_0 \sin(\alpha - \theta)$ [$M_0 \sin(\alpha' - \theta')$]. On the other hand, the transition matrix element of a minor transition is proportional to $\eta M_0 \cos(\alpha - \theta)$ [$\eta M_0 \cos(\alpha' - \theta')$], where $\eta$ depends on the incident laser energy. If we approximate the transition energy as $\varepsilon \cong v_0 p$, $\eta \cong \frac{\gamma_1}{2\varepsilon}[1+(1-\frac{\gamma_1}{\varepsilon})^2]$ for $\pi_1 \to \pi_2^*$, and $\eta \cong \frac{\gamma_1}{2\varepsilon}[1+(1+\frac{\gamma_1}{\varepsilon})^2]$ for $\pi_2 \to \pi_1^*$ transition. The Raman scattering intensity for the case in which only major transitions are involved can be approximated as

$$I_{major}(\alpha, \alpha') = I_o \left| \iint_0^{2\pi} \sin(\alpha-\theta) \sin^4(\frac{\theta-\theta'}{2}) \sin(\alpha'-\theta') d\theta d\theta' \right|^2, \quad (2)$$

where $\alpha$ is the incident polarization, $\alpha'$ the scattered polarization, and $I_0 = I_{major}(\alpha - \alpha' = 0)$. Because the inter-layer force constant is much smaller than the intra-layer one, the phonon dispersion and the intra-layer motions of the iTO-phonons of bilayer graphene are almost same as that of single-layer graphene. So the electron-phonon scattering matrix element between two electrons are taken to be proportional to $\sin^2[(\theta-\theta')/2]$, with a same functional form as that of single-layer graphene.[25] For a Raman scattering pathway involving both major and minor transitions, the intensity can be expressed as

$$I_{minor}(\alpha, \alpha') = I_o \left| \iint_0^{2\pi} \sin(\alpha-\theta) \sin^4(\frac{\theta-\theta'}{2}) \eta \cos(\alpha'-\theta') d\theta d\theta' \right|^2. \quad (3)$$

Because one of the transitions (photon absorption or emission) is induced by the minor transition, the $\sin(\alpha'-\theta')$ term in Eq. (2) is replaced with $\eta \cos(\alpha'-\theta')$. Contributions of Raman scattering processes involving only minor transitions are much smaller and can be ignored. Considering a projection of the scattered light to the analyzer with a polarization $\hat{e}_A = (\cos\varphi, \sin\varphi, 0)$, the resulting intensity after integrating out all relevant angles can be expressed as

$$I(\beta) = \int_0^{2\pi} (I_{major}(\alpha,\alpha') + I_{minor}(\alpha,\alpha'))\cos^2(\varphi-\alpha')d\alpha'$$
$$= \frac{I_{//}}{3+\eta^2}(1+2\cos^2\beta + \eta^2(3-2\cos^2\beta)), \qquad (4)$$

where $\beta = \varphi - \alpha$ is the relative angle between the incident polarization and the analyzer direction. From Eq. (4), we can see that the polarization ratio is $R = \frac{3+\eta^2}{1+3\eta^2}$. Fig. 5(c) shows the calculated polarization dependences of the major and the minor 2D Raman scatterings for an excitation energy of 1.96 eV. In this case, $\eta$ is about 0.19. Because the polarization dependence of the major and minor Raman peaks are opposite to each other, the ratio $R$ is smaller than the single-layer value of 3. The solid curve in Fig. 3(a) shows the calculated dependence of the ratio on the excitation energy, which reproduces the general trend of the experimental data. In this calculation, we assumed the same electron-phonon interaction for different scattering paths, which results in the same $R$ value for all $P_{ij}$'s. Here we note that our model does not consider the trigonal warping effect and that the validity of the model is limited to the excitation energy range of 1.4eV<ε<2.1eV which is confirmed by our *ab initio* calculations (not shown here).

## 4. Conclusion

We performed polarized Raman scattering measurements on Bernal-stacked bilayer graphene with 4 different excitation energies (1.96, 2.41, 2.54 and 2.81 eV). The overall polarization dependence is similar to that of single-layer graphene: the G band is isotropic whereas the 2D band shows a strong polarization dependence, being maximum when the polarizations of the incident and scattered photons are parallel and minimum when they are orthogonal to each other. However, the polarization ratio $R$ of the 2D band shows a dependence on the excitation energy, unlike the case of single-layer graphene. This is a result of contributions from minor transitions that have been usually ignored in previous analyses of the Raman 2D scattering. This demonstrates the importance of the details of electron-phonon and electron-photon interactions in understanding the Raman scattering processes in bilayer graphene.

**Acknowledgement**

This work was supported by the National Research Foundation (NRF) grants funded by the Korean government (MEST) (No. 2011-0013461 and No. 2011-0017605). This work was also supported by a grant (Nos. 2011-0031630 and 2011-0031640) from the center for Advanced Soft Electronics under the Global Frontier Research Program of MEST. Y.-W.S. was also supported by the NRF of Korea grant funded by the MEST (QMMRC, No. R11- 2008-053-01002-0). Computations were supported by the CAC of KIAS.